Title: The Occurrence Rate of Earth Analog Planets Orbiting Sunlike Stars

Authors: Joseph Catanzarite and Michael Shao, Jet Propulsion Laboratory, California Institute of Technology

**Abstract**
Kepler is a space telescope that searches Sun-like stars for planets. Its major goal is to determine $\eta_{Earth}$, the fraction of Sunlike stars that have planets like Earth. When a planet 'transits' or moves in front of a star, Kepler can measure the concomitant dimming of the starlight. From analysis of the first four months of those measurements for over 150,000 stars, Kepler's science team has determined sizes, surface temperatures, orbit sizes and periods for over a thousand new planet candidates. In this paper, we characterize the period probability distribution function of the super-Earth and Neptune planet candidates with periods up to 132 days, and find three distinct period regimes. For candidates with periods below 3 days the density increases sharply with increasing period; for periods between 3 and 30 days the density rises more gradually with increasing period, and for periods longer than 30 days, the density drops gradually with increasing period. We estimate that **1% to 3%** of stars like the Sun are expected to have Earth analog planets, based on the Kepler data release of Feb 2011. This estimate of $\eta_{Earth}$ is based on extrapolation from a fiducial subsample of the Kepler planet candidates that we chose to be nominally 'complete' (i.e., no missed detections) to the realm of the Earth-like planets, by means of simple power law models. The accuracy of the extrapolation will improve as more data from the Kepler mission is folded in. Accurate knowledge of $\eta_{Earth}$ is essential for the planning of future missions that will image and take spectra of Earthlike planets. Our result that Earths are relatively scarce means that a substantial effort will be needed to identify suitable target stars prior to these future missions.

**Methods**
The habitable zone (HZ) is conventionally defined as the region near a star where liquid water could exist at a planet's surface. The HZ limits for the solar system were determined to be $0.95\ AU < a < 1.37\ AU$ (Kasting, Whitmire, & Reynolds, 1993). However, in the Exoplanet Task Force (ExoPTF) Report (Lunine, 2008) it is pointed out that the reflecting properties of clouds could enable a planet to maintain a habitable surface temperature at closer orbit distances, and greenhouse gases such as $CO_2$ could make a planet warmer at larger orbit distances; neither of these effects were accounted for in the earlier study. Accordingly, the ExoPTF adopted a more optimistic HZ range of $0.75\ AU < a < 1.8\ AU$ for the solar system. If we define the 'scaled semimajor axis' as $s = \frac{a}{\sqrt{L}}$, where $L$ is luminosity in solar units, then a planet is in the habitable zone if either $0.95\ AU < s < 1.37\ AU$ or $0.75\ AU < a < 1.8\ AU$. If a planet is the in the HZ and its radius $r$ (in units of $R_{Earth}$) is within the Earth to super-Earth regime $0.8 < r < 2$, we define the planet as an Earth analog (EA). These intervals in $s$ and $r$ define the *Earth analog region* (see Figure 7). We adopt $2R_{Earth}$ as the maximum radius of a super-Earth (Borucki, 2011). The minimum radius of an Earth is $0.8R_{Earth}$ which corresponds to a mass of $0.5M_{Earth}$, the estimated minimum mass of a planet that could hold an Oxygen atmosphere.

We first count the number of *detected* transits in the FID region. The detected transits include false detections, so we estimate and subtract the number of these to determine $N_{transits,FID}$, the expected number of transiting planets among the detections in the FID region. We model the Kepler planet candidates by fitting power-laws for $r$ and $s$, and we use the models to analytically determine the ratio $M$ of transiting planets in the EA and FID regions. The number of transiting Earth analog planets is then $N_{transits,EA} = M \times N_{transits,FID}$. A transit can occur only if a planet's orbit plane is nearly aligned with the line of sight. For every star hosting a transiting planet, there are likely many other stars bearing planets with the same $r$ and $s$ whose orbit planes are not aligned favorably for transit. The probability of a transit for a planet orbiting a star of radius $R_{star}$ at semimajor axis $a$ and impact parameter smaller than $b$ is $P_{transit} = \frac{bR_{star}}{a}$. Here, impact parameter $b$ is defined as the perpendicular distance from the transit chord to the center of the star, in units of the star's radius, so $b$ ranges from 0 to 1. To account for the alignment effect we apply a geometric multiplier $A = \langle P_{transit}^{-1} \rangle = \langle \frac{a}{bR_{star}} \rangle$. We estimate the *total* number of Earth analog planets in the Kepler field as $N_{EA} = A \times N_{transits,EA} = A \times M \times N_{transits,FID}$. Then $\eta_{Earth} = \frac{N_{EA}}{153196} = \frac{A \times M \times N_{transits,FID}}{153196}$, the ratio of $N_{EA}$ to the total number of FGK stars in the Kepler field. Full description of each step of the calculation is provided in the following sections.

The estimate of $\eta_{Earth}$, the fraction of FGK stars with Earth analog planets, rests on two assumptions. The first is that the Kepler transit detections in the February 2011 data release are complete in a certain subset of phase space that we call the *fiducial region* (see Figure 7). We specify our choice of the fiducial region and give arguments for its completeness below. To derive our estimate of $\eta_{Earth}$, we extrapolated from the FID region to the EA region, by means of simple power laws in mass and scaled semimajor axis. So the second assumption is that the fitted mass and scaled semimajor axis power laws are valid in the contiguous region of phase space bounded on the left and right (in Figure 7) by the minimum scaled semimajor axis in the FID region and the maximum scaled semimajor axis in the EA region, and on top and bottom by the maximum planet radius in the FID region and the minimum planet radius in the EA region. If this is not the case, the extrapolation is invalid; in other words, the usual extrapolation caveat applies. If future Kepler data shows that the transit detections in the FID region were not complete, then the estimate for $\eta_{Earth}$ will most likely have to be revised upward.

**Selection criteria for the sample of transiting planet candidates**
The February 2011 Kepler data release contains 1235 candidate planets detected in the first ~4 months of science operations (Borucki, 2011). If a transit was detected during the first 132 days, all subsequent transits over the period Q0 – Q5 (about 400 days) were also used to refine the transit parameters included in the data release. Therefore detections of transiting planets with periods shorter than 132 days are complete as long as transit duration and depth provide sufficient SNR. Following the practice of the Kepler Science Team, we exclude the 17 planet candidates whose radius exceeds twice that of Jupiter as they are likely to be grazing binary companions. We also exclude candidates with a single observed transit (over the Kepler mission from Q0 – Q5) as these have a high likelihood of being false detections. Excluding the 33 candidates with both these attributes leaves 1202 transit candidates.

**The fiducial data set is nominally complete**
Correction for completeness is essential when calculating the fraction of stars that have planets. Kepler doesn't detect any planets of radius 0.1 $R_{Earth}$, but that doesn't mean they don't exist. We chose a fiducial sample with radii ranging from 2 to 4 $R_{Earth}$ and scaled semi-major axis ranging from 0.25 to 0.5 AU because the sample within those limits is nominally complete. We need to consider three types of completeness. The first is orbital period completeness. The sampling window must be long enough for Kepler to have seen at least two transits. The second is orbital inclination completeness. Only near edge-on orbits produce transits. Clearly the list of candidate planets includes only those with orbital inclinations near zero. The correction for orbital inclination completeness is outlined in the 'Methods' section above. The last is SNR completeness; a transiting planet must have SNR > 7 in order to be detected by Kepler.

The properties of the planet candidates announced by the Kepler project were determined from data spanning about 400 consecutive days during Q0 - Q5. Each detected planet candidate had at least one transit in the first 132 days. All transiting planets with period shorter than 132 days were detected, provided they had sufficient SNR .

The Kepler mission was originally built for the photometric precision to be 20 parts per million (ppm) for a 6.5 hour integration on a 12$^{th}$ magnitude star, with 10 ppm allocated to stellar variability and 17 ppm to photon noise from the star. After launch it was found that the average stellar variability is larger, ~25 ppm. The photometric precision of the Kepler spacecraft is given as 30 ppm (parts per million) for a 6.5 hour integration on a 12$^{th}$ magnitude star (Caldwell, 2010):

$$\sigma_{total}(K_p = 12, t = 6.5) = \sqrt{(17\ ppm)^2 + (25\ ppm)^2} = 30\ ppm.$$

If a star with Kepler magnitude $K_p$ has a planet transiting with duration $t$ in hours, then the photon noise per transit is

$$\sigma_{photon}(K_p, t) = 17\ ppm\ \sqrt{\left(\frac{6.5\ hours}{t}\right) \times 10^{\left(\frac{K_p - 12}{2.5}\right)}}\ .$$

If $N$ is the number of transits with duration $t$ in the 400 day survey, then the total noise in the transit measurements is

$$\sigma_{total}(K_p, t, N) = \sqrt{N^{-1} \times \left[\left(\sigma_{photon}(K_p, t)\right)^2 + (25\ ppm)^2\right]}.$$

The transit signal is $\left(\frac{R_{planet}}{R_{Sun}}\right)^2$ in $ppm$ , so we have

$$SNR(K_p, t, N) = \frac{N^{1/2} \times \left(\frac{R_{planet}}{R_{Sun}}\right)^2}{\sqrt{\left(\sigma_{photon}(K_p, t)\right)^2 + (25\ ppm)^2}}$$

Using the above equation for photometric accuracy, and estimating the transit duration using the orbit period and Kepler's law for a solar-type star, we find $SNR = 7$ for a 15.85 mag star with a planet of $2R_{Earth}$ and $P = 132$ days. Thus the extreme lower right-hand corner of the FID region is complete for stars all the way to 15.85 mag. A $2R_{Earth}$ planet orbiting a 16$^{th}$ mag star has $SNR > 7$ for $P < 108$ days. The box in phase space between 108 and 132 days and 15.85$^{th}$ mag and 16$^{th}$ mag is roughly 50% complete. However, less than 5% of the Kepler stars have magnitudes in that range, and of these only about 5% would have planet periods in the range of 108 to 132 days, if the density of planets were uniform in $log\ P$. We conclude that the fiducial region is nominally 100% complete.

For an Earth-radius planet transiting a 16[th] mag solar-type star, the photon noise in a 6.5 hr transit is 107 ppm, about 4 times larger than the average stellar variability noise of 25 ppm. The 25 ppm stellar noise is a serious impediment for detecting planets of the same size as the Earth. It is not a major noise source for detecting planets that are twice the Earth's size, because the photometric signal is four times larger.

**Comparison with (Howard & al., 2011) analysis**

For our sample we chose to accept all planets with orbit periods < 132 days that were detected with $SNR > 7$ over the 400 days of Q0 - Q5 and. Our criteria are less restrictive than the criteria used in (Howard & al., 2011) The planets in our fiducial region ($2 < r < 4$ and $0.25 < s < 0.5$) will not be complete in the analysis in (Howard & al., 2011), since they chose to use only 90 days of data. The main differences in the two samples are summarized in Table 1 below.

Table 1: Selection criteria for sample of planet candidates

| **This study** | (Howard & al., 2011) |
|---|---|
| $SNR > 7$ in 400 days | $SNR > 10$ in 90 days |
| All Kepler stars | $4100K < T_{eff} < 6100K$ |
| No restriction on $\log(g)$ | $4.0 < \log(g) < 4.9$ |
| $P < 132$ days | $P < 50$ days |

In Figure 1, we compare the period probability density function (PDF) of our sample of planet candidates with that of (Howard & al., 2011). The log period bin boundaries chosen for this plot are the same as those in Figure 2 of (Howard & al., 2011). Each curve represents planet candidates with radii between 2 and 4 $R_{Earth}$. The samples are each corrected for completeness. Because the threshold for the (Howard & al., 2011) sample was higher ($SNR$ = 10 in 90 days), that sample is less complete for the smaller radius planets. But after the correction for completeness in (Howard & al., 2011), the two samples look essentially the same.

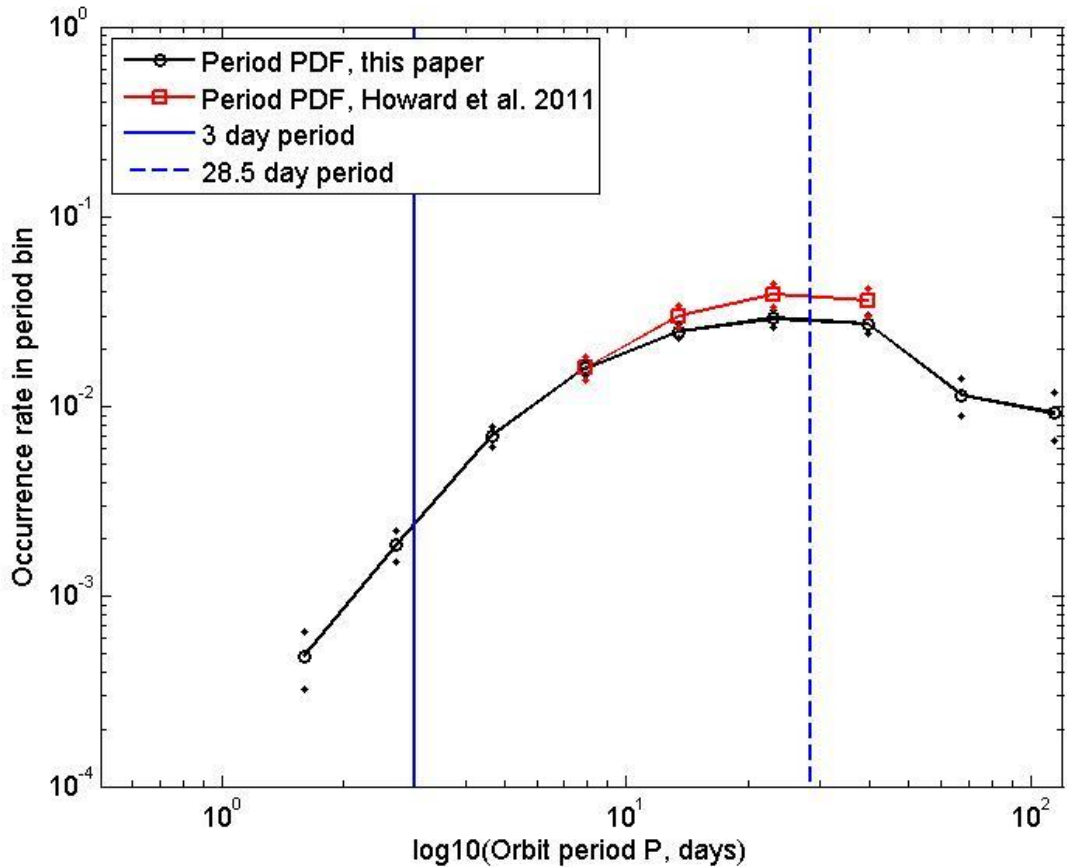

**Figure 1** Period distribution of planet candidates with radius between 2 and 4 $R_{Earth}$. The bin boundaries are the same as those in Fig 2 of (Howard & al., 2011). The samples are each corrected for completeness. Solid points indicate upper and lower Poisson error bars. The (Howard & al., 2011) sample goes up to 50 day period; our sample goes up to 132 day period.

The data presented this way has rather coarse bins in log period. Figure 2 shows the same data plotted with finer resolution. There are evidently three distinct power-law regimes for super-Earths and Neptunes. There is a sharp drop in the density of planets with periods < 3 days. Between 3 and 30 days, the density of planets increases with increasing period. But for periods between 30 and 132 days, the density of planets decreases with increasing period. Since our sample is nominally complete to periods of up to 132 days, we conclude that the period PDF changes from 'increasing with period' to 'decreasing with period' somewhere near P = 30 days, and that the period power-law fitted from (Howard & al., 2011) does not hold for periods beyond 50 days.

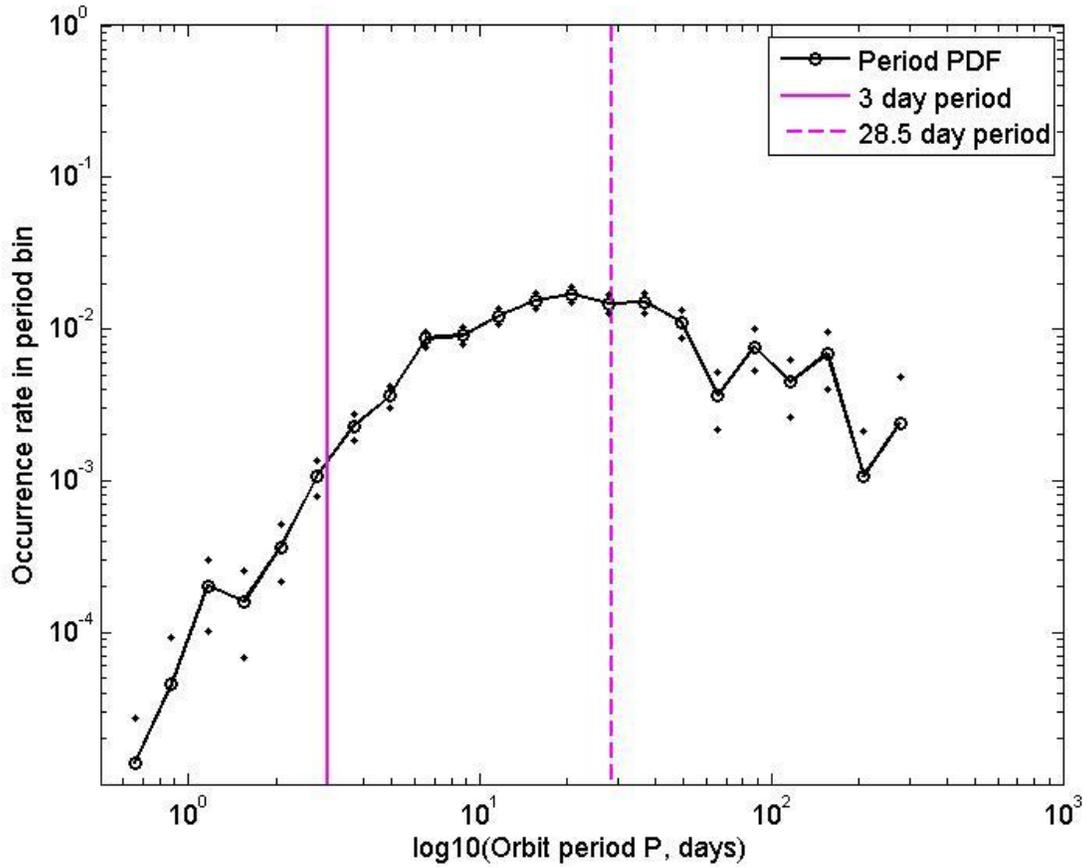

**Figure 2** Period distribution of planet candidates with radius between 2 and 4 $R_{Earth}$. The binning is finer than in Figure 1. Solid points indicate upper and lower Poisson error bars. The first two and last two bins each contain only 1 planet candidate, so their lower Poisson error bars are zero and therefore do not appear on the plot. A power-law was fitted to the data for periods greater than 28.5 days (see Figure 6). The last three bins contain planets with period longer than 132 days.

**Power-law fits**

We fitted the planet radius $r$, the scaled semimajor axis $s$, and the period P from the candidate planets to power-laws, employing the maximum-likelihood method (Clauset, Shalizi, & Newman, 2009). The technique fits unbinned data directly to a power-law and estimates the power-law index and the 'lower cutoff' of the data, i.e. the value below which the data is inconsistent with the power-law. If a continuous parameter $X$ has a power-law distribution, its cumulative distribution function (CDF) can be expressed as $CDF(x) \equiv P(X > x) = \left(\frac{x}{x_{min}}\right)^{-\alpha}$ where $\alpha$ is the power-law index, $x_{min}$ is the lower cutoff of the parameter, and $P(X > x)$ is the probability that the parameter $X$ takes on values larger than $x$. For ease of comparison with other studies of power-law distributions of planet parameters, we note that an alternative representation of the power-law is $\frac{dP(x)}{d(\log x)} = Cx^{-\alpha}$, where $C$ is a proportionality constant and $P(x)$ is the probability density function (PDF) of $x$.

The fitted planet radius power-law fit is shown in Figure 3. We find $CDF(r) = \left(\frac{r}{1.77}\right)^{-1.48}$. The planet radius power-law fit has a lower cutoff radius of $r_{min} = 1.77$; this is due to incompleteness when the planet is too small to be detected with $SNR > 7$. The fit is excellent out to $r \sim 5$. For comparison, Neptune has $r \sim 4$ and Jupiter has $r \sim 10$. For planets with radii $5 < r < 16$ the number density bulges above the fitted power-law; evidently, the power-law is not a good fit for these planets. Larger planets fall sharply away from the power-law. Such planets seem to follow a different power-law. It has been previously noted that this behavior is consistent with a prediction of accretion theory (Gould & Eastman, 2011). For our purpose, the behavior of the data for $r > 5$ is immaterial, since we will only use the fitted power-law in the interval $2 < r < 4$ to extrapolate number counts to the neighboring interval $0.8 < r < 2$.

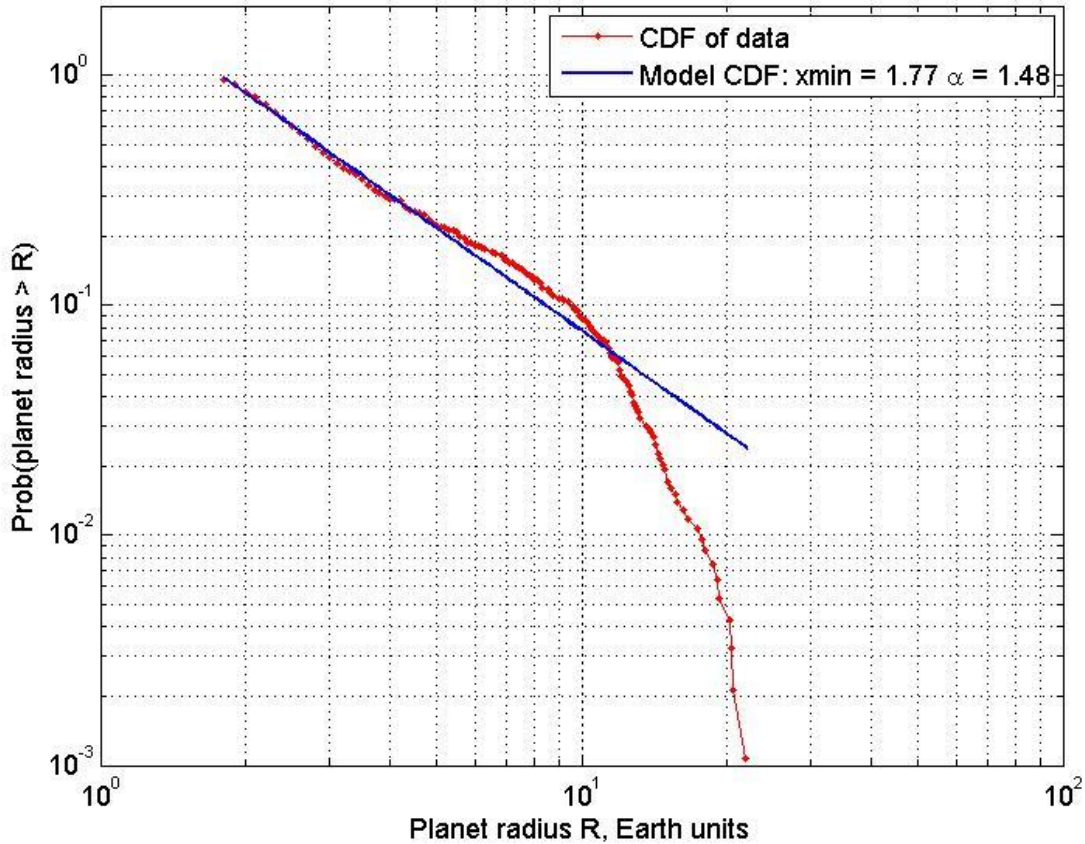

**Figure 3 Planet radius power-law model vs. cumulative distribution function for 1176 planet candidates with $P < 132$ days. The fitted power-law index is $\alpha = 1.48$. The planet radius power-law fit has a lower cutoff of $r_{min} = 1.77$; this is due to incompleteness when the planet is too small to be detected with sufficient SNR. The fit is excellent out to $r \sim 5$.**

The scaled semimajor axis power-law fit is shown in Figure 4. We find $CDF(s) = \left(\frac{s}{0.24}\right)^{-2.54}$. The power-law fit has a lower cutoff of $s_{min} = 0.24 \, AU$, corresponding to an orbit period of about 40 days for a G2 star. The power-law fit to the scaled semimajor axis $s$ is remarkably good. The data falls away from the model for $s > 1\ AU$, due to incompleteness in long-period candidates with at least two observed transits. The scaled semimajor axis $s$ corresponds to a unique equilibrium temperature that can be derived from the Stefan-Boltzmann radiation law, $T_{eq} = T_\odot (1-A)^{0.25} \sqrt{\frac{R_\odot}{2a}}$, where $a$ is the orbital semimajor axis, $A$ is the planet's albedo, and $T_\odot$ and $R_\odot$ are the Sun's temperature and radius.

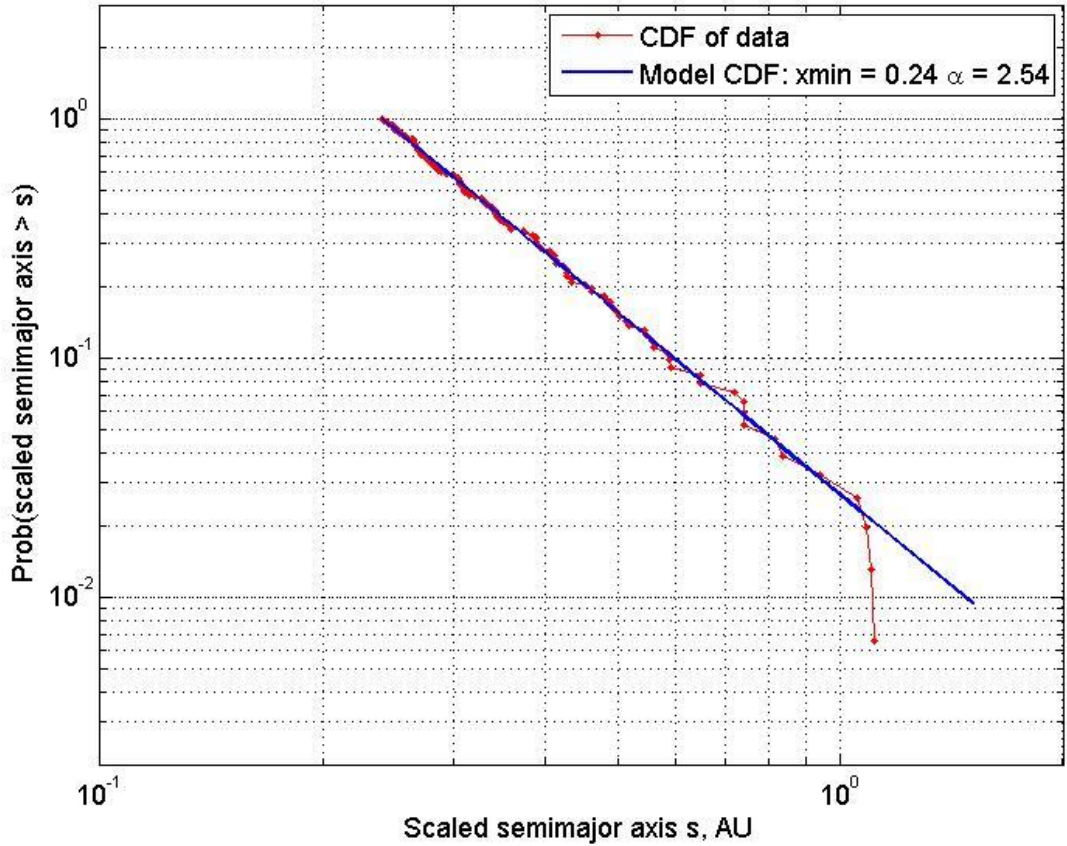

**Figure 4** Scaled semimajor axis power-law model vs. cumulative distribution function for 577 Kepler planet candidates with $< 4$. The fitted semimajor axis power-law index is $\alpha = 2.54$ The scaled semimajor axis power-law fit has a lower cutoff of $s_{min} = 0.24\ AU$, corresponding to an orbit period of about 40 days for a G2 star. The data falls away from the model for $s > 1\ AU$ due to incompleteness in long-period candidates with at least two observed transits. Care must be taken in comparing power-law fits in semimajor axis $a$ for transit-selected planets to RV surveys, because the sensitivity to transits varies as $a^{-1}$. To compare our semimajor axis fits with those derived from RV surveys we account for this selection effect by subtracting 1 from the fitted power-law index. The semimajor axis power-law index is $\alpha = 1.54$ (after this correction).

It is evident from Figures 3 and 4 that power-law distributions provide excellent fits to the Kepler data in the fiducial region in the $(r, s)$ phase space defined by $2 < r < 4$ and $0.25\ AU < s < 0.5\ AU$. Kepler detected 87 transiting planet candidates with impact parameter $b < 0.85$ in the fiducial region.

The period power-law fit is shown in Figure 6. The Feb 2010 Kepler data release should include every detectable planet with orbit period of less than 132 days (Borucki, 2011).

As of the February 2011 data release, the Kepler pipeline could not automatically include data spanning multiple quarters in its transit fits. Planets with a period longer than 90 days either had enough SNR to be seen in a single transit, or data from multiple 90 day data sets had to be manually stitched together. It might be conjectured that the February 2011 Kepler transit discoveries should therefore be significantly incomplete at periods longer than 90 days. However, detection of a single transit doesn't require SNR of 7. For a planet to be listed in the Feb 2011 list of 1235 planets, the ~400 day SNR must be > 7.

In our analysis we assume the planet list is nominally complete for R> 2Rearth and P<132 days. We tested this by comparing planet occurrence rates vs. period for faint ( Kp<14) vs. bright (Kp>14 ) stars. Faint stars are ~2 mag dimmer than bright stars, so their SNR is lower by a factor of 2.5. If their detection completeness were significantly lower than that of bright stars, we would expect the occurrence rates of longer period planets to be much smaller for faint stars than for bright stars. As shown in Figure 5 below, this is not the case. Occurrence rates of faint stars are about 1-sigma lower at long periods, but this may not be statistically significant.

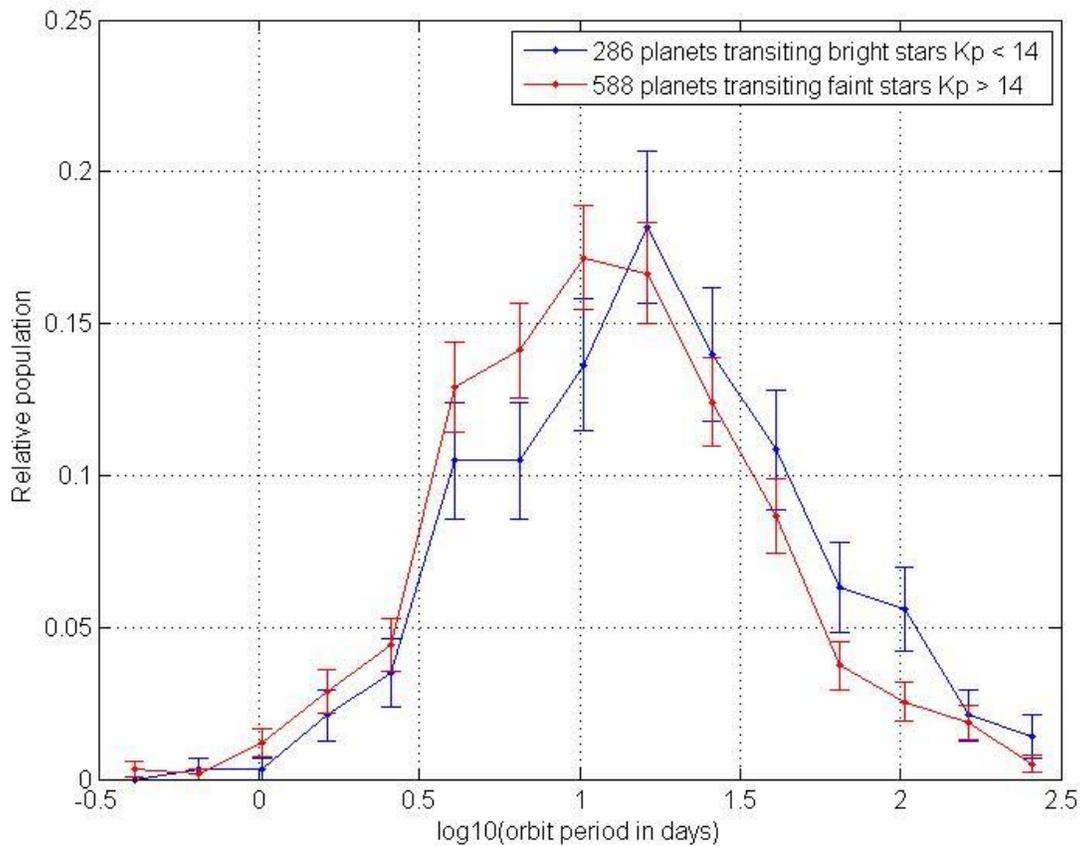

**Figure 5 Comparison of detected population of planets with R>2 vs. orbit period, for bright and faint stars. Bins are 0.2 dex in log(P).**

The period power law fit is shown in Figure 6. The fitted power-law index is $\alpha = 1.87$. Evidently the power-law provides a reasonable fit to the data for periods between 30 and 130 days; the planet density at longer periods falls off due to incompleteness. The lower cutoff for period is 28.5 days, which means that planets with shorter periods do not fit this power law. The reason is evident from Figure 2, which shows that the period PDF is divided into three distinct regimes. The distribution function drops sharply for planets with periods shorter than about 3 to 6 days, perhaps due to some physical mechanism which depletes their population. Care must be taken in comparing power-law fits in semimajor axis $a$ or period $P$ for transit-selected planets to RV surveys because the sensitivity to transits varies as $a^{-1}$ and therefore as $P^{-2/3}$. We correct for this selection effect by subtracting $\frac{2}{3}$ from the fitted power-law index. The corrected period power-law index is $\alpha = 1.20$.

Interestingly, we find that the density of planets decreases with period for periods longer than about 30 days. This differs markedly with the findings of previous RV surveys of Saturns and Jupiters (Tabachnik & Tremaine, 2002); (Cumming, Butler, Marcy, Vogt, Wright, & Fischer, 2008), which predicted an increase in the density of planets toward 1 AU. These RV surveys did not include super-Earth and Neptune planets, which dominate the Kepler planet candidates.

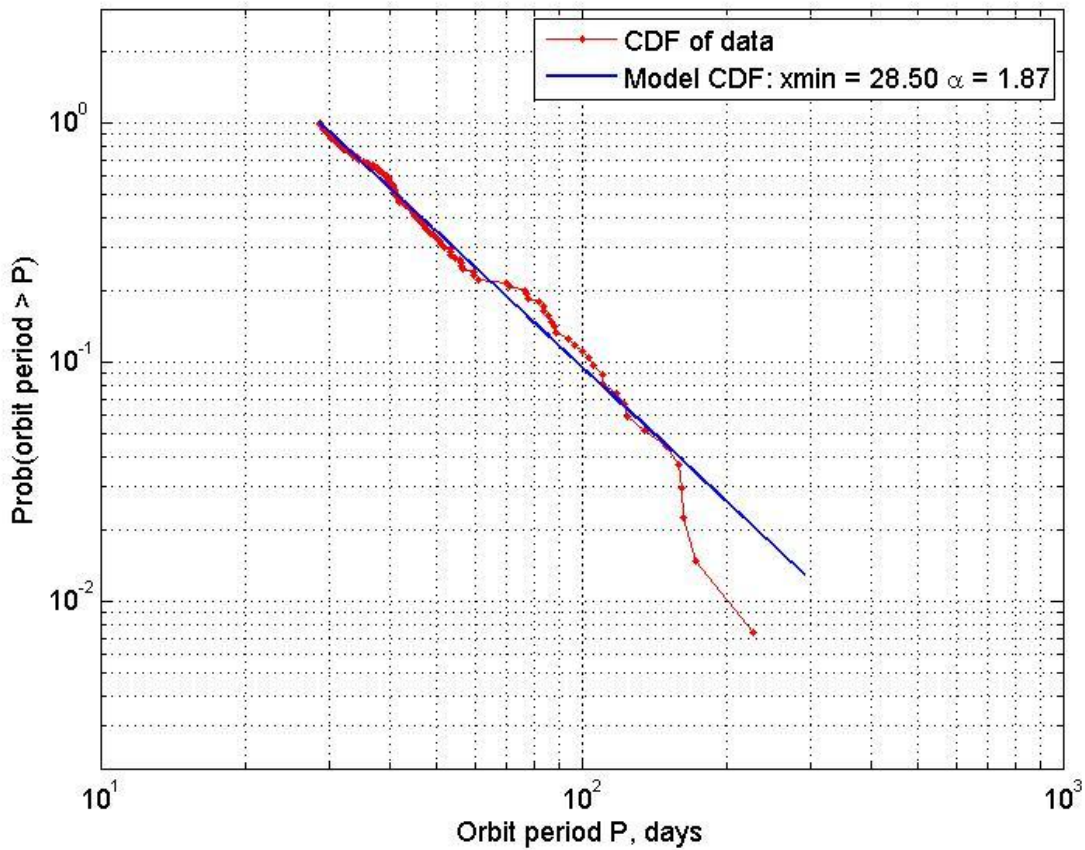

Figure 6 Orbit period power-law model vs. cumulative distribution function for 577 Kepler planet candidates with $r < 4$. The fitted period power law index is $\alpha = 1.87$. The lower cutoff for period is 28.5 days. Shorter period planets fall below the fitted power-law (see Figure 2), perhaps due to some physical mechanism which depletes their population. Care must be taken in comparing power-law fits in semimajor axis $a$ or period $P$ for transit-selected planets to RV surveys because the sensitivity to transits varies as $a^{-1}$ and therefore as $P^{-2/3}$. To compare our period fits with those derived from RV surveys we account for this selection effect by subtracting $\frac{2}{3}$ from the fitted power-law index. The corrected period power-law index is $\alpha = 1.20$. Interestingly, we find that the density of planets decreases toward longer periods. This differs markedly with the findings of previous RV surveys of Saturns and Jupiters (Tabachnik & Tremaine, 2002); (Cumming, Butler, Marcy, Vogt, Wright, & Fischer, 2008), which predicted an increase in the density of planets toward 1 AU. These RV surveys did not include super-Earth and Neptune planets, which dominate the planet candidates found by Kepler.

Care must be taken when using fitted power-laws, as the errors on the indices introduce uncertainty. The 1-sigma errors in the fitted power-law indices are given by $\sigma_\alpha = \frac{\hat{\alpha}-1}{\sqrt{N}}$, where $\hat{\alpha}$ is the estimated power-law index and $N$ is the number of measurements used to determine $\hat{\alpha}$.

Table 2 presents the fitted power-law indices and their upper and lower 1-sigma bounds.

**Table 2 Fitted power-law indices and their errors**

| Parameter | Lower bound $\alpha - \sigma_\alpha$ | Fitted index $\alpha$ | Upper bound $\alpha - \sigma_\alpha$ |
|---|---|---|---|
| **Planet radius, $r$** | 1.43 | 1.48 | 1.53 |
| **Scaled semimajor axis, $s$** | 2.33 | 2.54 | 2.74 |
| **Orbital period, $P$** | 1.71 | 1.87 | 2.03 |

**Correction for false alarms**

Though the data includes all transits with $b < 0.85$ in the FID region, it also includes a substantial number of false detections. The Kepler Science Team assigned each candidate a 'vetting flag' that is associated with its likelihood of being a false detection. The candidates fall into three categories, with false detection probabilities of 2%, 20%, 40%. Subsequent analysis showed that candidates whose host stars are brighter than 14 magnitudes have only 10% chance of being false detections (Morton & Johnson, 2011). We therefore re-assigned stars brighter than 14[th] magnitude to a fourth category with false detection probability of 10%. For the candidates in the Kepler data set, the weighted mean false alarm probability is 23%. We used a Monte Carlo approach to determine the expected number of false transit detections among the fiducial candidates. For each of 1000 survey realizations, we removed 2%, 20%, 40%, and 10% of the detected candidates in the four respective categories from the data before fitting the power-laws. The power-law indices for $r$ and $s$ were fairly robust to the removal of the ensemble of simulated false detections; they changed only slightly. The number of transits in the fiducial region is $N_{transits,FID} = 65.4 \pm 3.6$ after accounting for false detections.

**Number of transiting Earth analogs in the Kepler field**

If we adopt the first definition of the HZ, a planet is in the HZ if $0.95\ AU < s < 1.37\ AU$ (Kasting, Whitmire, & Reynolds, 1993). A plot of planet radius vs. scaled semimajor axis for the Kepler planet candidates is shown in Figure6. The 'Earth analog' (EA) and 'Fiducial' (FID) regions are bordered by the red and cyan boxes, respectively. The centers of the EA and FID regions are $\sim 0.5\ dex$ apart in $s$ and $0.35\ dex$ apart in $r$.

The next step is to use the fitted power laws in $s$ and $r$ together with the number of transiting planets detected by Kepler in the FID region to extrapolate the number of transiting planets in the EA region. We note that extrapolation of the power laws is correct only if there is no new physics in the formation of planets that would dominate in the area outside where there is complete data. However, the Kepler release does contain planets with periods longer than 130 days, and radii smaller than $2R_{Earth}$, and though this data is incomplete, there is no obvious evidence that the power laws break down in the phase space regime extending from the FID region to the EA region. With this caveat, we assume that the power-law fits in $s$ and $r$ extend from the FID region into the EA region, and we compute the ratio $M$ of planet candidates in these regions by integrating the power-law models:

$$M = \frac{\int_{0.8}^{2} \frac{dP(r)}{dr} dr \int_{0.95}^{1.37} \frac{dP(s)}{ds} ds}{\int_{2}^{4} \frac{dP(r)}{dr} dr \int_{0.25}^{0.5} \frac{dP(s)}{ds} ds} = 0.104_{-0.027}^{+0.050}$$

The error bars are from the uncertainty in the fitted power-law indices. To determine them, we re-ran the calculation of $\eta_{Earth}$ using the upper and lower 1-sigma bounds on the power-law indices given in Table 2.

Next we use $M$ to extrapolate the number of transiting Earth analogs expected in the Kepler field of view from the number of transiting planets detected in the fiducial region.

$$N_{transits,EA} = M \times N_{transits,FID} = 6.8_{-1.8}^{+3.3}$$

The uncertainty in $N_{transits,EA}$ comes from errors in $M$ and in $N_{transits,FID}$.

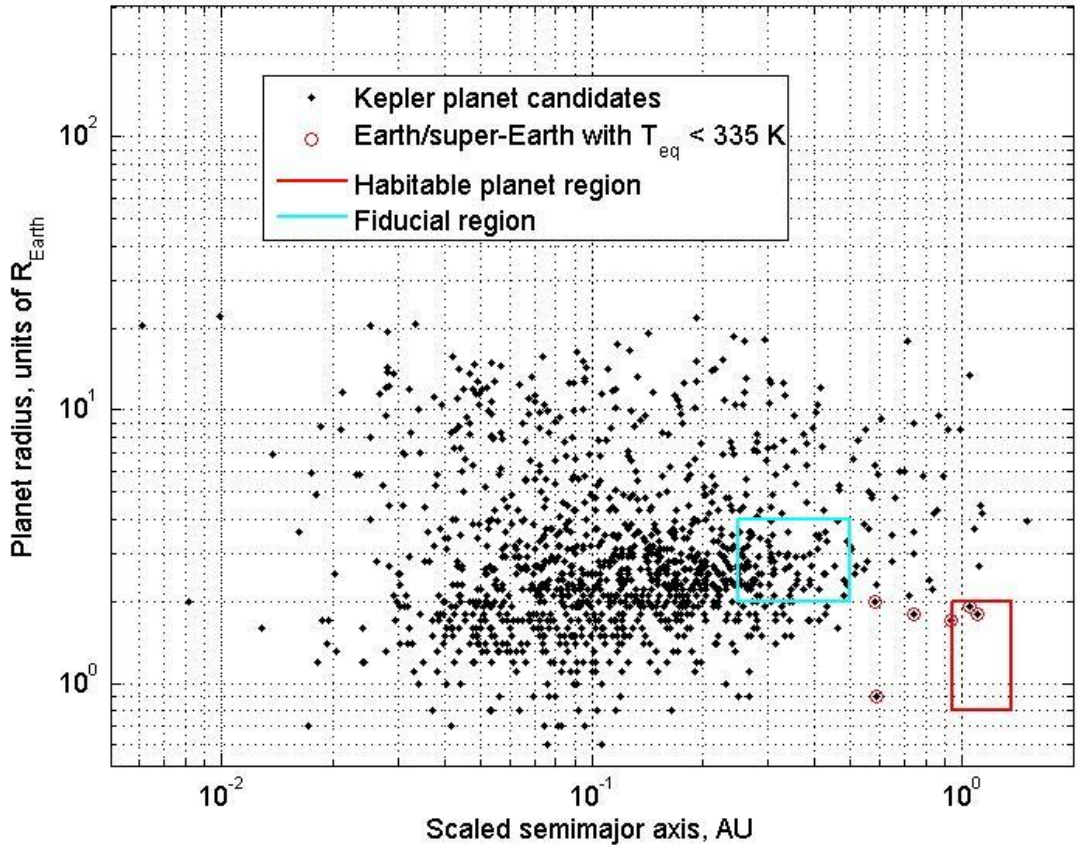

**Figure 7** Kepler transit candidates represented by black points in the phase space of scaled semimajor axis vs. planet radius. Scaled semimajor axis is $s \equiv \frac{a}{\sqrt{L}}$, where $a$ is the semimajor axis of the planet's orbit in AU and $L$ is the luminosity of the host star, in solar units. The Earth analog region $0.95 < s < 1.37$ and $0.8 < r < 2$ is within the red box, and the fiducial region $0.25 < s < 0.5$ and $2 < r < 4$ is within the cyan box. Six Kepler candidates with $r < 2$ that are in or near the habitable

zone are circled in red. Extrapolation from the center of the fiducial region to the center of the Earth analog region spans $\sim 0.5\ dex$ in $s$ and $\sim 0.35\ dex$ in $r$.

**Correction for geometric alignment**

Kepler detects only transiting planets, whose orbit planes are aligned or nearly aligned with the observer's line of sight. The inclination range for transits is $\frac{\pi}{2} - \frac{b_{max}R_{star}}{a} < i < \frac{\pi}{2} + \frac{b_{max}R_{star}}{a}$, where $a$ is the orbital semi major axis, $i$ is the orbital inclination and $b_{max}$ is the largest impact parameter for which a transit can be reliably detected. We find that in the fiducial region, the distribution of $b$ for planet candidates has a plateau between 0.65 and 0.85 and falls off rapidly at larger $b$; only 3 planet candidates were detected with $b > 0.85$. We adopt a value of $b_{max} = 0.85$ and assume that all transiting planets in the FID region with smaller impact parameter will be detected. From Monte Carlo simulations, we found that varying the choice of $b_{max}$ introduces ~2% error in the geometric correction.

Orbit plane inclinations are randomly distributed with probability density $\frac{dP(i)}{di} = 0.5 \sin i$. Integrating the density over the inclination range we find that the probability of a transit is $P_{transit} = 0.85 \frac{R_{star}}{a}$, where $R_{star}$ is the radius of the star, and $a$ is the planet's orbital semimajor axis. Therefore, for every transiting planet with semimajor axis $a$ orbiting a star with radius $R_{star}$ there are $\frac{1}{P_{transit}} = \frac{a}{0.85 R_{star}}$ planets with the same semimajor axis, with random orbit inclinations.

To estimate the alignment correction factor $A = \langle \frac{a}{0.85 R_{star}} \rangle$ we randomly drew an ensemble of 100,000 planets from the fitted power-law distributions for $r$ and $s$ with scaled semimajor axis and radius within the fiducial region. For each planet we randomly drew a host star from the candidate host star data provided by the Kepler Science Team, and recorded its effective temperature $T_{eff}$ and stellar radius $R_{star}$. The luminosity of a star is $L = R_{star}^2 \left(\frac{T_{eff}}{T_{sun}}\right)^4$ from the Stefan-Boltzmann law. In terms of the scaled semimajor axis $s$, the planet's semimajor axis is $a = s\sqrt{L}$. We found a geometric alignment correction factor with mean and standard deviation $A = 255 \pm 64$ for the ensemble of 100,000 simulated planets in the EA region. Correcting for geometric alignment, and for false detections, we find the expected number of Earth analog planets in the Kepler field is $N_{EA} = A \times N_{EA,transits} = A \times M \times N_{FID,transits} = 1732 \pm 322$.

**Calculation of $\eta_{Earth}$**

The February 2011 data release contains data for 153,196 stars in the Kepler field. We therefore estimate that the fraction of habitable planets for all Sun-like stars is $\boldsymbol{\eta_{Earth}} = \frac{N_{EA}}{N_{stars}} = \mathbf{1.1^{+0.6}_{-0.3}\%}$

For the preceding calculation of $\boldsymbol{\eta_{Earth}}$ we had adopted the HZ range $0.95\ AU < s < 1.37\ AU$ (Kasting, Whitmire, & Reynolds, 1993). We note that $\eta_{Earth}$ is sensitive to the choice of habitable zone boundaries. If we adopt the HZ boundaries $0.75\ AU < s < 1.8\ AU$ from the ExoPTF report (Lunine, 2008), we estimate $\boldsymbol{\eta_{Earth}} = \mathbf{2.8^{+1.9}_{-0.9}\%}$. Taking both estimates together, we find that the most likely range of $\boldsymbol{\eta_{Earth}}$ is between 1.1% to 2.8%, with 1-$\sigma$ minimum of 0.8% and 1-$\sigma$ maximum of 4.7%

Results with both HZ definitions are summarized in Table 3 below.

Table 3 Range of $\eta_{Earth}$ for Kasting and ExoPTF HZ boundaries

| HZ definition | **Lower bound of $\eta_{Earth}$** | **Estimated $\eta_{Earth}$** | **Upper bound of $\eta_{Earth}$** |
|---|---|---|---|
| Kasting (1993) | 0.8% | 1.1% | 1.7% |
| ExoPTF (2008) | 1.9% | 2.8% | 4.7% |

**The problem of Subgiants**

Kepler stars that are hotter than $5400\ K$ present a problem, because the algorithm to estimate their surface gravity is not effective. As a result, every star hotter than $5400\ K$ is assigned the surface gravity of a main sequence star at the same temperature. This means that the radii of subgiants hotter than $5400\ K$ are underestimated by a factor of 1.5 to 2. (Brown, Latham, Everett, & Esquerdo, 2011). A glance at the plot of $T_{eff}$ vs. $R_{star}$ in Figure 8 shows a narrow spiky region just to the left of $T_{eff} = 5400\ K$ containing stars with $R_{star}$ between 2 and 4. These are the cool subgiants for which meaningful stellar radii *can* be estimated. This population does not extend beyond $T_{eff} = 5400\ K$; possibly the estimated radii of hotter subgiants are systematically biased downward so that they masquerade as main sequence stars. If a transit candidate is a subgiant hotter than $T_{eff} = 5400\ K$, its radius and orbital semimajor axis will also be underestimated by the same factor of 1.5 to 2 because these are derived by multiplying the measurements $\frac{a}{R_{star}}$ and $\frac{R_{planet}}{R_{star}}$ by $R_{star}$. If there are many subgiants among the transit candidates, the power-law fit for r could potentially be compromised. However, the power-law fit for s will *not* be compromised, because $s = \frac{a}{\sqrt{L}} = \frac{a}{R_{star}T_{eff}^2}$, and the measurement $\frac{a}{R_{star}}$ is not affected by the bias in the apriori value of $R_{star}$. In any case, inspection of Figure 8 shows that there are at most 2 or 3 subgiants among the fiducial stars that are cooler than $5400\ K$, indicating that cool subgiants are not so numerous among the Kepler candidates. As a check, we excluded the stars that are hotter than $5400\ K$ from the fiducial data set, and used only the remaining cool stars to determine an estimate $\eta_{Earth,cool\ stars} = 0.009$ based *only* on stars cooler than $5400\ K$, for the Kasting HZ boundaries $0.95\ AU < s < 1.37\ AU$. We correct this to include the hotter stars (assuming they contribute approximately the same fraction of transits) by the multiplier $\frac{N_{cool}+N_{hot}}{N_{cool}}$ to get $\eta_{Earth} = 0.016$. Since this result is within the range we determined for the Kasting HZ using all the stars in the fiducial region, we conclude that hot subgiants do not cause an appreciable error in our determination of $\eta_{Earth}$.

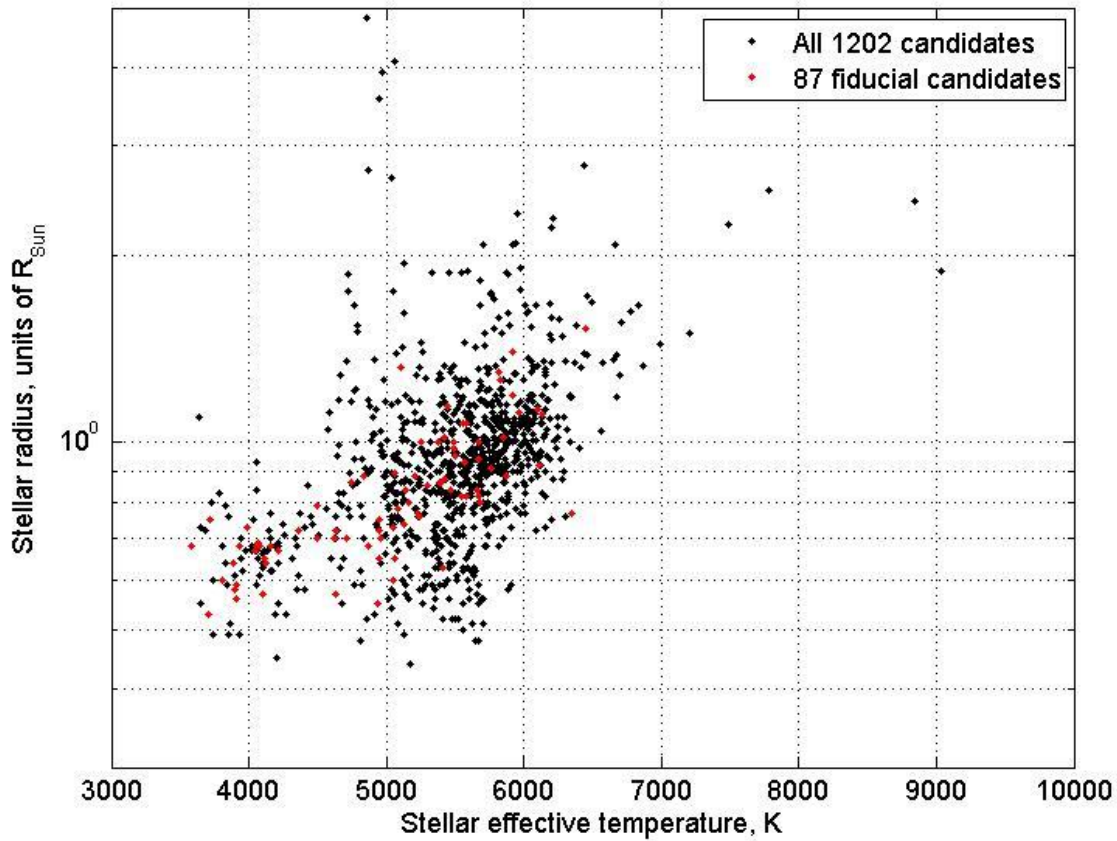

**Figure 8 Stellar effective temperature vs. radius**
**Kepler stars that are hotter than 5400 $K$ present a problem, because the algorithm to estimate their surface gravity is not effective. As a result, every star hotter than 5400 $K$ is assigned the surface gravity of a main sequence star at the same temperature. This means that the radii of subgiants hotter than 5400 $K$ are underestimated by a factor of 1.5 to 2.** (Brown, Latham, Everett, & Esquerdo, 2011). **Note the narrow spiky region to the left of $T_{eff} = 5400\ K$ (log ($T_{eff}$) =3.73) containing stars with $R_{star}$ between 2 and 4 times the Sun's radius. These are the cool subgiants for which meaningful stellar radii *can* be estimated. Evidently the radii of hotter subgiants are pushed down to masquerade as main sequence stars.**

**Comparison with previous studies**
Previous investigators have analyzed the masses and periods of planets discovered by radial velocity surveys (Tabachnik & Tremaine, 2002); (Cumming, Butler, Marcy, Vogt, Wright, & Fischer, 2008). In the first study the mass and period distributions of a sample of Jupiter-mass planets were modeled as power-laws. A small but significant correlation was found in mass and period distributions. The second study extended the work into the Saturn-mass regime and periods out to 2000 days. Most recently (Howard & al., 2010) used a complete sample of planets discovered by the radial velocity (RV) method with orbit periods shorter than 50 days and masses down to super-Earths from the Eta Earth Survey to characterize the mass distribution of close-in super-Earths and Neptunes.. This distribution was then extrapolated to predict the occurrence rate of Earth-mass planets. They found that 23% of FGK stars have planets with masses from $0.5 - 2 M_{Earth}$ with periods shorter than 50 days. In all these studies, a power-law is represented as $\frac{dN}{dlogx} \sim x^{-\alpha}$. Note that the index $\alpha$ in our power-law model $CDF(x) =$ $\left(\frac{x}{x_{min}}\right)^{-\alpha}$ is the same as the index in the alternate representation $\frac{dN}{dlogx} \sim x^{-\alpha}$. Our fit to the period power-law for Kepler transit candidates is shown in Figure 6. The lower cutoff for period is 28.5 days; shorter period planets may follow a different power-law, as is evident from Figures 1 and 2. Care must be taken in interpreting fitted power-law indices in semimajor axis $a$ or period $p$ for transit-selected planets, because the sensitivity to transits varies as $a^{-1}$ and therefore as $p^{-2/3}$. To compare our period fits with those derived from RV surveys we can account for this selection effect by subtracting $\frac{2}{3}$ from the fitted power-law index. The Eta Earth Survey estimated the mass distribution of Neptunes and super-Earths by fitting planet masses to a power-law. Kepler measures planet radius, not mass. Nevertheless, we can estimate the planet power-law index from our fitted radius power-law index by assuming that planet mass $m$ is proportional to $r^3$ in the Earth/super-Earth regime. Then, since $\frac{dN}{dlogm} = \frac{dN}{dlogr}\frac{dlogr}{dlogm} \sim \frac{dN}{dlogr} \sim r^{-\alpha_r} \sim m^{-\alpha_r/3}$, the planet mass and radius power-law indices are related by $\alpha_m = \alpha_r/3$. In Table 3, we compare the mass and period power-law fits from the Kepler survey to those of RV surveys. The fitted period power law index from our study is corrected for the transit selection effect by subtracting $\frac{2}{3}$, as discussed in this section.

Table 3 Mass and period power-law indices derived from the Kepler transit survey compared to previous RV surveys

| Survey | Index $\alpha$ for power-law $\frac{dN}{dlogx} \sim x^{-\alpha}$ | | $\alpha_m$ (planet mass) |
|---|---|---|---|
| | $\alpha_p$ (orbit period) | | |
| RV (Tabachnik & Tremaine) | $-0.27 \pm 0.06$ | | $0.11 \pm 0.10$ |
| RV (Cumming et al.) | $-0.26 \pm 0.1$ | | $0.31 \pm 0.2$ |
| RV (Howard et al., 2010) | N/A | | $0.48 \pm 0.14$ |
| Kepler transits (corrected for selection effect) | P = 30 to 130 days, $2 < r < 4$, this work | P = 3 to 30 days, $2 < r < 4$, Howard et al., 2011 | $0.49 \pm 0.02$ |
| | $+1.20 \pm 0.16$ | $-0.27 \pm 0.27$ | |

The Feb 2011 Kepler data probes super-Earth and Neptune mass planets nominally complete out to periods of ~130 days and significantly incomplete out to periods of ~300 days. The Eta-Earth survey probes a similar mass range in orbits closer than 50 days. It is notable that the fitted mass power-laws in the last three rows are in fairly good agreement, suggesting that the close-in planets come from the same population as those in the more distant orbits. On the other hand, the period distribution power-law for the Kepler transit candidates is very different from that determined by the first two RV surveys. However there is no reason to expect the power-laws to be the same, since these RV surveys probe gas giants of Saturn and Jupiter mass, while the planet candidates being found by the Kepler survey are mostly super-Earths and Neptunes.

Our new finding from analysis of the Kepler data shown in Figure 2 is that a single power-law cannot be used to describe the planet density function of super-Earths and Neptunes with log (period). There are three distinct period regimes. For periods shorter than 3 days, the density of planets drops precipitously with decreasing period. Planet density rises with increasing periods between 3 and 30 days, and it falls with increasing period for periods longer than ~30 days. Kepler reported planets with periods out to almost 300 days, and the cumulative period probability distribution is shown in Figure 6. At periods longer than ~130 days, the sample is significantly incomplete, and the number of planets is small, so the statistical error is large. But we do not see evidence in this data of an increase in the planet density (vs. log period) with longer periods.

Our estimate of $\eta_{Earth}$ derives from an extrapolation based on a power-law fitted to the planet candidates in the third region. The validity of this power law rests on our assumption that the transit detections are complete in this region. Had we had ignored planets with periods longer than 40 days in our analysis, the fitted period power law would have had a positive slope, since the planet density increases with period for periods less than 40 days (see Figure 2). Our extrapolated estimate of $\eta_{Earth}$ would accordingly have been much larger.

**Summary and conclusion**

We present a calculation of $\eta_{Earth}$, the occurrence rate of Earth analog planets orbiting FGK stars, based on the February 2011 Kepler data release. $\eta_{Earth}$ depends on the adopted definition of the HZ. For the conventional HZ boundaries (Kasting, Whitmire, & Reynolds, 1993), we find $\boldsymbol{\eta_{Earth} = 1.1^{+0.6}_{-0.3}\%}$. For the less conservative HZ boundaries given in the ExoPTF Report, we find $\boldsymbol{\eta_{Earth} = 2.8^{+1.9}_{-0.9}\%}$. Our Earth would not have liquid water on its surface at either 0.75 AU or at 1.8 AU. But some type of terrestrial planet could have liquid water closer to the Sun and a terrestrial planet with more green house warming might have liquid water near 2AU.

We base our estimate of $\eta_{Earth}$ on extrapolation from a fiducial region (in which the transit detections are complete) to the Earth analog region (in which they are significantly incomplete), using fitted power-law models of the radius and scaled semimajor axis distributions. We have implicitly assumed that the fitted mass and scaled semimajor axis power laws can be extrapolated from the fiducial region to the Earth Analog region in Figure 7. The caveat is that if these simple models do not extend all the way to the Earth Analog region, then the extrapolation is not valid. However, because of Kepler's sensitivity to small planets, this is the most robust extrapolation of planet distribution functions to Earth analog planets that can be made at this time.

We fitted power-laws to the period, planet radius and scaled semimajor axis distributions and compared our results to previous surveys. We estimated the mass power-law index by using a simple cubic scaling relation between mass and radius. The result is in excellent agreement with that of the Eta Earth Survey **(Howard & al., 2010)**, for super-Earths and Neptunes in orbits shorter than 50 days. Interestingly, we find that the density of super-Earth and Neptune planets decreases toward longer periods. This differs markedly with the findings of previous RV surveys of Saturns and Jupiters **(Tabachnik & Tremaine, 2002)**; **(Cumming, Butler, Marcy, Vogt, Wright, & Fischer, 2008)**, which predicted an increase in the density of planets toward 1 AU. However, these RV surveys included only gas giants, and did not include super-Earth and Neptune planets, which dominate the planet candidates found by Kepler.

$\boldsymbol{\eta_{Earth}}$ is a key parameter for planning future space missions to directly image and measure the spectra of exo-Earths. Many mission concepts that have been studied could potentially succeed only if $\boldsymbol{\eta_{Earth}}$ is large, at least 20% **(Savransky, Kasdin, & Cady, 2010)**; **(Catanzarite & Shao, 2011)**. A significantly smaller value of $\boldsymbol{\eta_{Earth}}$ means that a mission to detect nearby Earths would have to be capable of searching 100 or more of the nearest stars instead of 10 or 15. The 2010 Astrophysics Decadal Review **(New Worlds, New Horizons in Astronomy and Astrophysics, 2010)** stated that "the role of target-finding for future direct detection missions (is) not universally accepted as essential." But if $\boldsymbol{\eta_{Earth} = 2\%}$, substantial effort may be needed to identify suitable target stars prior to these future missions, possibly using space astrometry **(Zhai, Shao, Goullioud, & Nemati, 2011)**, **(Shao, Catanzarite, & Pan, 2010)**. With an estimate of $\boldsymbol{\eta_{Earth}}$ in hand we can improve the fidelity of the science modeling for these missions, which will make it possible to rank them by their relative merits.

**Acknowledgements**

This work was carried out at the Jet Propulsion Laboratory, California Institute of Technology, under contract with NASA. Copyright 2011 California Institute of Technology. Government sponsorship acknowledged. We thank Wes Traub for useful discusssions. JC thanks Stuart Shaklan for asking how the Kepler data constrain $\eta_{Earth}$.